# TIMBRE LATENT SPACE: EXPLORATION AND CREATIVE ASPECTS

Antoine CAILLON[1], Adrien BITTON[1], Brice GATINET, Philippe ESLING

Institut de Recherche et Coordination Acoustique Musique (IRCAM)
UPMC - CNRS UMR 9912 - 1, Place Igor Stravinsky, F-75004 Paris
{caillon, bitton, gatinet, esling}@ircam.fr

## Introduction

Recent studies show the ability of unsupervised models to learn invertible audio representations using Auto-Encoders (Engel et al., 2017). While they allow high quality sound synthesis and high-level representation learning, the dimensionality of the latent space and the lack of interpretability of each dimension preclude their intuitive use. The emergence of disentangled representations was studied in Variational Auto-Encoders (VAEs) (Kingma et al., 2014, Higgins et al., 2017) and has been applied to audio. Using an additional perceptual regularization (Esling et al., 2018) can align such latent representation with the previously established multi-dimensional timbre spaces, while allowing continuous inference and synthesis. Alternatively, some specific sound attributes can be learned as control variables (Bitton et al., 2019) while unsupervised dimensions account for the remaining features. In this paper, we propose two models and suited interfaces that were developed in collaboration with music composers in order to explore the potential of VAEs for creative sound manipulations[2]. Besides sharing a common analysis and synthesis structure, one has a continuous latent representation and another has a discrete representation, which are applied to learning and controlling loudness invariant sound features.

## Models

We consider a dataset of audio samples, such as performance recordings of an instrument. A variable-length audio $x$ can be processed by analyzing series $\{x_0, ..., x_L\}$ of signal windows $x_i \in R^{d_x}$ with an encoder $E_\phi$ mapping each frame into a latent code as $E_\phi : x_i \mapsto z_i \in R^{d_z}$. This encoder is paired with a decoder $D_\theta$ that inverts these features as $D_\theta : z_i \mapsto \hat{x}_i$. The vanilla auto-encoder optimizes its parameters $\{\theta, \phi\}$ on a reconstruction objective such that $\hat{x} \approx x$ (Figure 1).

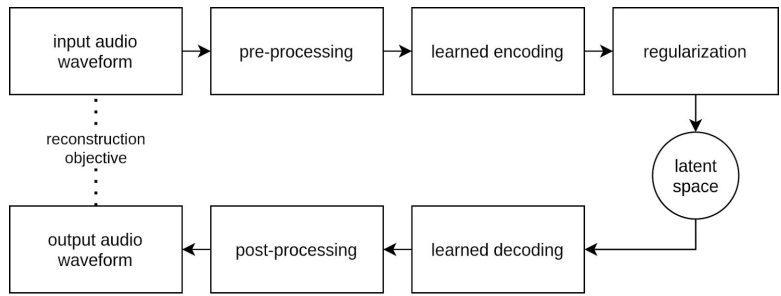

Figure 1. Block diagram of a VAE with optional pre and post audio processing.

Usually, we choose $d_z \ll d_x$ so that the latent variables embed a compressed representation of the data from which we can synthesize new samples. However, this continuous representation often remains highly-dimensional and does not disentangle data properties on separate latent dimensions. The usability of such representation and its quality for sampling or interpolation are thus limited. These considerations highlight the need for additional training objectives that enforce useful properties in the latent representation. We consider two separate models, comparable in their overall encoder-decoder structure[3], but different in how the representation is regularized during training.

---

[1] These authors contributed equally
[2] See https://acids-ircam.github.io/timbre_exploration/ for additional information about models, interfaces and sound examples.

*Continuous model.* The first model aims to construct a latent space that is invariant to loudness in order to embed features that mainly account for the instrument timbre. It is achieved with an adversarial domain adaptation, where a latent regressor is trained at predicting loudness, and a gradient reversal optimization (Ganin et al., 2015) leads to a loudness-invariant encoder representation. Besides this objective, the VAE latent space is regularized on a Gaussian prior distribution n $N(0, 1)$ which ensures local smoothness and favors independence between latent variables.

*Discrete model.* The second model is based on the Vector-Quantized VAE (VQ-VAE) proposed in van den Oord et al. (2017). It optimizes a discrete set of latent features $q^j$. Each encoder output is matched to its nearest codebook element $q_i^* \in \{q^0, ..., q^k\}$, before being decoded. This latent space is disentangled from a gain applied to the decoder output, which produces short-term features that are invariant to audio levels. Given that the set of latent features $q^j$ is finite, we can analyze and map this codebook with acoustic descriptors.

Both models are intended to learn latent audio features that are invariant to loudness. The continuous model offers unconstrained and smooth feature manipulations. The discrete model can be analyzed in order to predict the output acoustic features embedded in the representation.

**Experiments**

*Descriptor-based synthesis.* Each vector of the discrete representation is individually decoded and the output signal is analyzed with a descriptor. It is thus possible to compute the mapping between a descriptor curve and the series of nearest latent features (details in Bitton et al., 2020). Latent synthesis can be directly controlled by following a user-defined descriptor target, as shown in figure 2. The codebook can be ordered and traversed according to different properties, such as centroid or fundamental frequency.

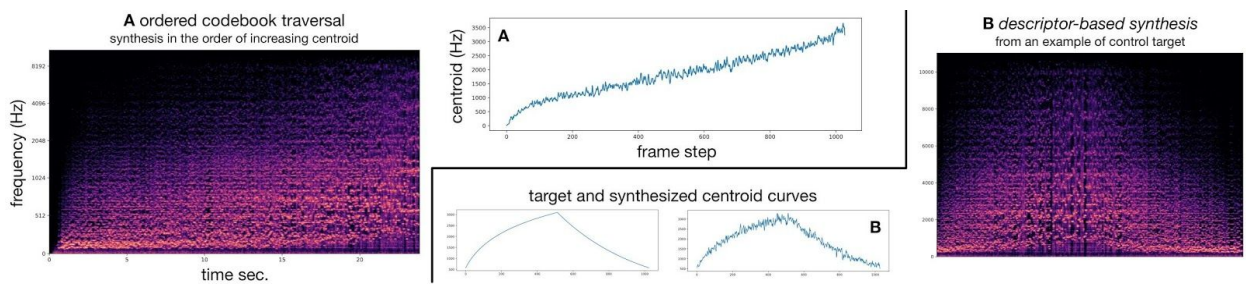

*Figure 2. The discrete representation can be analyzed with the spectral centroid and traversed in the increasing order (**A**). A control target can be synthesized by selecting the nearest latent features, the decoded audio approximately follows the curve provided (**B**).*

*Continuous latent interpolations.* In order to display the local smoothness of the continuous model, we consider the time variant linear interpolation $z_{interp}$ between two latent series

---

[3] Architectural differences are not detailed in this paper since we focus on discussing the representation properties.

$z_a$ and $z_b$ of the same size inferred from two audio samples A and B. Decoding $z_{interp}$ results in an audio sample smoothly interpolating between sample A and sample B, as shown in figure 3.

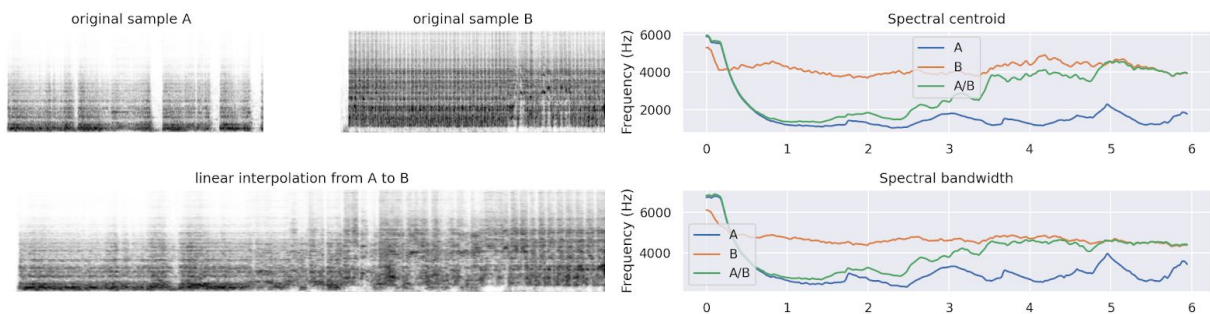

*Figure 3. Linear interpolation in the latent space between two audio samples. We can see that the centroid and the bandwidth of the interpolated audio sample performs a smooth transition between those of the two original audios.*

In order to facilitate a creative use of this model, we present two interfaces designed to circumvent the problem of identifying latent dimensions by facilitating their exploration.

**Continuous model interfaces**

The first interface is a Max/MSP application that is a graphical equivalent[4] to the command line tools we usually have to test the model. It features several high-level interactors such as mathematical operators on the latent series, manual editing, and an interpolation plane. We have built this application in collaboration with A. Schubert[5], aligning with his remarks on how to improve visualization and control over the generation. This interface is intended to be used in order to grasp the main characteristics of a model trained on a specific dataset.

This stand-alone interface has built-in interactions but a limited integration and restrictions in the possible operations. We have thus developed a second interface built in collaboration with B. Gatinet, implementing the encoder and the decoder as PureData abstractions that can be combined with any other regular objects. New aspects of the continuous model emerge from this interface, as it allows uninterrupted exploration with realtime rendering, enabling the use of complex signal processing techniques on both the audio and latent series. As this interface can be integrated in real time inside a digital audio workstation, it is more suited for composition workflows. It is furthermore a strict superset of the first interface in terms of functionalities.

The use of these interfaces has brought to light new ways of generating audio signals, whether by explicit control of an audio descriptor, or by morphing between different existing sounds. Training a model on an audio domain and using it to resynthesize an audio sample from a different domain can also lead to an implicit synthesis method. Additional results on audio conversion of instrument sounds can be found in Bitton et al., (2020).

**Conclusion**

This research has studied VAEs with continuous and discrete latent sound representations as creative tools to explore timbre synthesis. The discrete model allows the generation of a new audio signal by directly controlling acoustic descriptors. Manipulations of the continuous model are eased by developing specific interfaces and real-time rendering, which greatly enrich composition and sound design possibilities. And in turn, it gives further insights on the generative qualities found in the learned representations, as well as the relevance of their different parameters and controls with respect to the new timbres that are synthesized.

---

[4] See our website for a screenshot of the interface
[5] See http://www.alexanderschubert.net/

## Acknowledgment

This work is supported by the ANR:17-CE38-0015-01 MAKIMOno project, the SSHRC:895-2018-1023 ACTOR Partnership and Emergence(s) ACIDITEAM project from Ville de Paris and ACIMO projet from Sorbonne Université. The authors would also like to thank Alexander Schubert for his creative inputs.